# Quasi-confined modes produced by the Lugiato-Lefever model with a localized pump and the pseudo-Raman term

E. M. Gromov[1, *] and B. A. Malomed[2,3]

[1]National Research University Higher School of Economics, Nizhny Novgorod 603155, Russia

[2]Department of Physical Electronics, School of Electrical Engineering, Faculty of Engineering, and Center for Light-Matter Interaction, Tel Aviv University, Tel Aviv 69978, Israel

[3]Instituto de Alta Investigación, Universidad de Tarapacá, Casilla 7D, Arica, Chile

**Abstract**

We introduce an extended nonlinear Lugiato-Lefever equation (LLE) with the *pseudo-stimulated-Raman-scattering* (pseudo-SRS) cubic term, linear damping/gain, and spatial inhomogeneous (weekly or strongly localized) pump. The LLE is derived, in the extended adiabatic approximation, from the underlying Zakharov's system (ZS), which includes a viscosity term acting on its low-frequency (LF) component and the pump supporting the high-frequency (HF) one. Dynamics of quasi-solitons in the model is addressed by means of analytical and numerical methods. The sech-based approximation for the quasi-soliton predicts it as a stable fixed point (FP) of the system of evolution equations for the system's moments (the HF norm, wave momentum, and center-of-mass coordinate). The FP's attraction basin is identified too. The prediction is corroborated by direct simulations of the full LLE. A quasi-singular mode, supported by the delta-functional pump, is briefly considered too.

**Highlights**

>Quasi-localized states of an extended Lugiato-Lefever equation (LLE) are considered. >LLE includes the pseudo-stimulated-Raman-scattering effect and localized external pump. >The LLE is derived from the Zakharov's system. >Dynamics of the solitons is investigated by means of analytical and numerical methods. >Some modes supported by singular pumps are produced too.

## 1. Introduction

---

* Corresponding author. *E-mail address:* egromov@hse.ru (E.M. Gromov).

Solitons are robust localized modes which are maintained in a great variety of physical realizations by the stable balance of nonlinearity and linear dispersion or diffraction [1-4]. These realizations and the corresponding models are commonly categorized according to their basic features. In particular, in many cases they may be classified as high-frequency (HF) and low-frequency (LF) systems. In the former case, solitons appear as broad self-trapped envelopes of HF oscillations, well-known examples being provided by solitons in optical media with the Kerr nonlinearity [1-3] (which may include left-handed metamaterials [5-9]), Langmuir solitons in plasmas [10-12], matter-wave solitons in ultracold bosonic gases [13-16], nonlinear surface-wave excitations on deep water [17-19], and Davydov solitons in long molecules [20,21]. A universal model for envelope solitons is provided by the nonlinear Schrödinger equation (NLSE) [22,23]. On the other hand, nonlinear LF fields may directly self-trap into solitons, such as ones built by ion-acoustic [24] and magneto-acoustic [25] waves in plasmas, and surface waves on shallow water layers and long internal waves in stratified fluids [26,27]. The fundamental model for nonlinear LF media is given by the Korteweg – de Vries equation (KdVE) and its generalizations.

More complex systems often include interactions of HF and LF waves. The interest to such systems was originally drawn by the Zakharov's system (ZS) [17], which was derived for the nonlinear coupling between Langmuir (HF) and ion-acoustic (LF) waves in plasmas. Unlike the above-mentioned NLSE and KdVE in their basic forms, the ZS is not an integrable model [28]. Nevertheless, it gives rise to stable single-soliton solutions. Other well-elaborated implementations of the ZS represent the interplay of HF oscillations of atomic positions and acoustic waves in long molecular chains which may realize Davydov solitons [29], and the coupling of HF surface and LF internal waves in stratified fluids [30].

The ZS may be simplified, replacing the second-order bidirectional LF equation by one reduced for the unidirectional propagation. In particular, KdVE may naturally play the role of the latter equation, being coupled to the NLSE for HF waves, see Eqs. (1) and (2) below. The LF component may be further eliminated in the adiabatic approximation, in which the slowly evolving LF field is replaced by a term representing the local density of the HF component. In the adiabatic approximation, the ZS reduces to the NLSE.

In many cases a realistic model includes viscous damping of the LF wave. In the presence of the damping, the adiabatic approximation still makes it possible to replace to the full ZS by the single-component NLSE, which includes a dissipative nonlinear term, see the one with coefficient $\mu$ below in Eq. (3). The same term in the NLSE with time $t$ and coordinate $x$ replaced, respectively, by the propagation distance, $z$, and "retarded time", $\tau$, represents the effect of the stimulated Raman scattering (SRS) in fiber optics [31-33], i.e., splitting of the pump photon into a longer-wavelength secondary one (the *Stokes wave*) and an optical phonon (the *idler*, which represents an effective loss, in terms of optical fields). Phenomenologically, the SRS effect in fibers is a manifestation of a small delay (~a few fs) of the nonlinear response of silica to the temporal variation of the optical field [1,3,32]. On the contrary to the time-domain SRS in fiber optics, the formally similar term in Eq. (3) emulates the SRS effect in the spatial domain. Therefore, it was called a *pseudo-SRS* effect in works [34-38], where it was derived, as outlined above, by means of the adiabatic approximation applied to the ZS including viscosity acting in the LF component. The pseudo-SRS effect was studied in the dynamics of solitons [34-37], as well as in terms of the modulational instability of continuous waves [38]. A common feature of the temporal-domain SRS and spatial pseudo-SRS effects is their dissipative nature.

The dynamics of narrow temporal solitons under the action of the SRS effect in fiber optics was studied in detail experimentally [31] and theoretically [32,33,1-3]. In particular, it is well known that SRS shifts the soliton's spectrum towards longer wavelengths, which may be interpreted as an effective self-deceleration of the temporal solitons, which eventually leads to their destruction. This trend may be compensated by the use of a frequency-sliding gain [39], which is applied to compensate losses in long-haul fiber-optic communication lines [39]. Other possibilities for the suppression of the SRS-induced drift of the spectrum are offered by the emission of small-amplitude waves from the temporal soliton [40], periodic modulation of the group-velocity dispersion [41,42], shift of the zero-dispersion point [43], and the use of dispersion-decreasing fibers [44]. In addition to inducing the self-deceleration of the solitons, SRS produces a similar effect on optical shock waves [45,46]

Possibilities for stabilization of spatial solitons against the action of the pseudo-SRS effect were analyzed too [34-37,47]. In particular, it was demonstrated that the stabilization may be provided by variable dispersion [34-37] or by an external potential, which may be naturally added to the effective NLSE in the contexts of the underlying physical realizations in the spatial domain [47].

Another relevant possibility, which was not addressed previously, is that the physical settings modeled by the NLSE with the pseudo-SRS term may also include an external drive in the form of a HF pump with a spatial inhomogeneous amplitude. In particular, in the original context of the ZS for plasmas, the pump may represent external time-modulated jets used to control the behavior of the hot plasma [48]. Similarly, in terms of the surface waves, the spatial inhomogeneous pump may represent external fluid jets interacting with the surface [49]. The objective of the present work is to introduce such a setup and consider the dynamics of one-dimensional localized states in it, by means of combined analytical and numerical methods. As explained in detail below, the same model may be realized as the Lugiato-Lefever equation (LLE), i.e., an externally pumped NLSE with linear loss [50-52], which in this case includes the SRS term, cf. Refs. [53-59]. Accordingly, the localized modes predicted in the present work may be realized as well in settings modeled by the LLE, which chiefly refer to laser cavities [60].

The subsequent presentation is organized as follows. The model is formulated in Section 2, and analytical results are reported in Section 3. Using evolution equations for the norm, wave momentum, and center-of-mass coordinate, applied to the sech ansatz, we predict a quasi-soliton solution as a fixed point (FP) of the respective dynamical system, along with its attraction basin. In Section 4, the predictions of the analytical approximation for the quasi-soliton are corroborated by direct simulations. That section also presents brief results for a quasi-soliton pinned to the singular pump in the form of the delta-function. The paper is concluded by Section 5.

## 2. The model and integral relations

We consider the evolution of a slowly varying envelope $U(x,t)$ of the HF wave field (e.g., the local amplitude of Langmuir waves in the plasma) in the one-dimensional medium with the cubic nonlinearity, under the action of linear loss/gain with the damping/amplification real rate $\nu$ and spatiotemporally-dependent external pump $f(x,t)$. The corresponding NLSE is nonlinearly coupled to a unidirectional equation for the LF field, $n(x,t)$ (e.g., local perturbation of the ion density of the plasma), including the viscosity/diffusion term with coefficient $\mu$. As a result, we arrive at the following generalized ZS:

$$2i\frac{\partial U}{\partial t}+\frac{\partial^2 U}{\partial x^2}-2nU+i\nu U = f(x,t), \qquad (1)$$

$$\frac{\partial n}{\partial t}+\frac{\partial n}{\partial x}-\mu\frac{\partial^2 n}{\partial x^2}=-\frac{\partial |U|^2}{\partial x}, \qquad (2)$$

where the coefficients in front of the second and fourth terms in Eq. (2) are fixed to be 1 by means of scaling. In the lowest adiabatic approximation, which neglects the derivatives $\partial n/\partial t$ and $\partial^2 n/\partial x^2$ in Eq. (2), reduces it to a local relation, $n = -|U|^2$. Accordingly, Eq. (1) is replaced by the damped/amplified driven NSLE:

$$2i\frac{\partial U}{\partial t}+\frac{\partial^2 U}{\partial x^2}+2U|U|^2+i\nu U = f(x,t). \qquad (3)$$

In the next approximation, the diffusion term in Eq. (2), which was neglected at the previous stage, is substituted by $\mu\partial^2 n/\partial x^2 \approx -\mu\partial^2(|U|^2)/\partial x^2$, to produce an amended local relation, $n = -|U|^2 - \mu\partial(|U|^2)/\partial x$ and, consequently, an extended NLSE:

$$2i\frac{\partial U}{\partial t}+\frac{\partial^2 U}{\partial x^2}+2U|U|^2+\mu U\frac{\partial\left(|U|^2\right)}{\partial x}+i\nu U=f(x,t). \tag{4}$$

Here the term $\sim \mu$ represents the above-mentioned pseudo-SRS effect in the spatial domain. We adopt the pump term as

$$f=|f(x)|\exp(i\phi-i\omega_0 t), \tag{5}$$

with frequency $\omega_0$ and phase shift $\phi$ (in terms of the usual form of LLE [50]), $\omega_0$ is tantamount to the mismatch parameter).

Note that, for the smoothly inhomogeneous pump (5), with a spatial scale $L$, which satisfies the constraint $\partial^2 f/\partial x^2 \sim f/L^2 \ll |\omega_0|\cdot|f|$, the small-amplitude background (bg) wave field far from the "body" of the soliton is

$$U=V_{bg}(x)\exp(i\phi-i\omega_0 t),\ V_{bg}(x)\approx |f(x)|/(2\omega_0+i\nu). \tag{6}$$

This result is valid under condition $|f(x)|^2 \ll 4|\omega_0|^3$, which makes it possible to neglect the effect of the cubic term on the background amplitude. Numerical results are presented below for values of parameters satisfying this condition.

As mention above, Eq. (3) is well known as LLE, which was originally introduced as the spatial-domain NLSE for the envelope of the electromagnetic field in driven lossy nonlinear cavities [50,60], and later extended, as a time-domain equation, for modeling the frequency-comb generation in dissipative Kerr resonators [51,52]. In the latter context, LLE may include higher-order dispersion, represented by linear terms with higher-order derivatives, added to Eq. (3) [51]. In the context of the frequency-comb generation, the LLE including the SRS term was introduced too, and dynamics of solitons was studied in that setting in detail, both theoretically and experimentally [53-59]. However, in those works the LLE (3) was studied with a constant (spatially uniform) pump, $f = \text{const}$, while our intention here is to consider NLSE (3) with the pseudo-SRS term and a spatiotemporally-modulated pump, $f(x,t)$. It is relevant to mention that localized states in LLE with a spatially localized pump, $f(x)$, were recently studied too [61-63]. This possibility is briefly addressed at the end of section 4.

We are looking for solutions to Eq. (4) in the form of

$$U(x,t)=V_{bg}(x)\exp(i\varphi-i\omega_0 t)+\Psi(x,t)\equiv V(x,t)+\Psi(x,t), \tag{7}$$

where the background term $V_{bg}(x)\exp(i\phi-i\omega_0 t)$ is the one defined by Eq. (6), and the proper soliton term is localized, so that $\Psi|_{x\to\pm\infty}\to 0$. Assuming that the soliton's width is much smaller than the scale $L$ of the spatial inhomogeneity of the pump term, we substitute ansatz (7) in Eq. (4) and omit the second spatial derivative of the background term, as adopted above, which yields the effective equation for the soliton field:

$$2i\frac{\partial \Psi}{\partial t}+\frac{\partial^2 \Psi}{\partial x^2}+2\Psi|\Psi|^2+4V(x,t)|\Psi|^2+2\Psi^2 V^*(x,t)+i\nu\Psi+\mu\Psi\frac{\partial(|\Psi|^2)}{\partial x}=0, \quad (8)$$

where * stands for the complex conjugate. With the intention to address quasi-solitons, we adopt for them zero boundary conditions in the infinite domain, $viz.$, $\Psi|_{x\to\pm\infty}\to 0$. Then, the following evolution equations for naturally defined integral moments of the HF field are easily derived from Eq. (8):

$$\frac{dN}{dt}\equiv\frac{d}{dt}\int_{-\infty}^{\infty}|\Psi|^2 dx=-\nu N+i\int_{-\infty}^{+\infty}|\Psi|^2\left(V(x,t)\Psi^*-V^*(x,t)\Psi\right)dx, \quad (9)$$

$$2\frac{dP}{dt}\equiv 2\frac{d}{dt}\int_{-\infty}^{\infty}k|\Psi|^2 dx=-\mu\int_{-\infty}^{+\infty}\left[\frac{\partial(|\Psi|^2)}{\partial x}\right]^2 dx-2\nu kN-$$

$$4\int_{-\infty}^{+\infty}|\Psi|^2\left(\frac{\partial\Psi}{\partial x}V^*(x,t)+\frac{\partial\Psi^*}{\partial x}V(x,t)\right)dx-$$

$$2\int_{-\infty}^{+\infty}\left(\frac{\partial\Psi^*}{\partial x}\Psi^2 V^*(x,t)+\frac{\partial\Psi}{\partial x}(\Psi^*)^2 V(x,t)\right)dx-$$

$$3\int_{-\infty}^{+\infty}|\Psi|^2\left(\frac{\partial V(x,t)}{\partial x}\Psi^*+\frac{\partial V^*(x,t)}{\partial x}\Psi\right)dx, \quad (10)$$

$$\frac{dM}{dt}\equiv\frac{d}{dt}\int_{-\infty}^{\infty}x|\Psi|^2 dx=-i\int_{-\infty}^{+\infty}\left(\Psi^*\frac{\partial\Psi}{\partial x}-\Psi\frac{\partial\Psi^*}{\partial x}\right)dx-\nu\int_{-\infty}^{\infty}x|\Psi|^2+$$

$$i\int_{-\infty}^{+\infty}x|\Psi|^2\left(V(x,t)\Psi^*-V^*(x,t)\Psi\right)dx. \quad (11)$$

Here, we define the Madelung representation $\Psi=|\Psi|\exp(i\vartheta)$, with $k=\partial\vartheta/\partial x$ being the local wavenumber. Obviously, $N$, $P$ and $M$ define the total norm and wave momentum of the wave packet, while $M$ is the product of the packet's center-of-mass coordinate and norm.

## 3. Analytical approximations
### 3.1. Effective evolution equations

To analyze of the wave-packet dynamics in the framework of Eqs. (9)-(11), we adopt the usual sech ansatz for solutions, with variable amplitude $A(t)$, wavenumber $k(t)$, and central coordinate $\bar{x}(t)$:

$$\Psi(x,t) = A(t)\mathrm{sech}[A(t)(x-\bar{x}(t))]\exp\left[ik(t)x + i(1/2)\int A^2(t)dt\right]. \tag{12}$$

The substitution of ansatz (12) and pump term (5) in Eqs. (9)-(11), using the above constraints that the spatial scale of the pump term is much larger than the soliton's width, and assuming that the losses/amplification rate is small, $|\nu| \ll |\omega_0|$, we derive the following system of evolution equations for the time-dependent wavenumber, amplitude, phase $\theta(t) \equiv (1/2)\int A^2(t)dt + \omega_0 t$, and central coordinate:

$$2\frac{dk}{dt} = -\frac{8}{15}\mu A^4 - \frac{|f(\bar{x})|Ak}{3\omega_0} I\left(\frac{k}{A}\right)\sin(\phi-\theta) - \frac{3A}{2\omega_0}\frac{d|f(\bar{x})|}{dx} I\left(\frac{k}{A}\right)\cos(\phi-\theta), \tag{13}$$

$$2\frac{dA}{dt} = -2\nu A - \frac{|f(\bar{x})|A^2}{\omega_0} I\left(\frac{k}{A}\right)\sin(\phi-\theta), \tag{14}$$

$$2\frac{d\theta}{dt} = \Omega \equiv A^2 + 2\omega_0, \quad \frac{d\bar{x}}{dt} = k, \tag{15}$$

where $d|f(\bar{x})|/dx \equiv (d|f(x)|/dx)_{x=\bar{x}}$ is the spatial gradient of the pump at the soliton's central point, and

$$I\left(\frac{k}{A}\right) \equiv \int_{-\infty}^{\infty} \frac{\cos(kz/A)}{\cosh^3 z} dz. \tag{16}$$

For further consideration we assume the natural assumption that the wavenumber of ansatz (10) is small in comparison with the soliton's inverse width, viz., $|k|/A \ll 1$, hence $I(k/A) \approx \pi/2$, in Eqs. (13), (14). Then Eqs. (13)-(15) simplify to

$$2\frac{dk}{dt} = -\frac{8}{15}\mu A^4 - \frac{\pi|f(\bar{x})|}{6\omega_0} Ak\sin(\phi-\theta) - \frac{3\pi}{4\omega_0}\frac{d|f(\bar{x})|}{dx} A\cos(\phi-\theta), \tag{17}$$

$$2\frac{dA}{dt} = -2\nu A - \frac{\pi|f(\bar{x})|}{2\omega_0} A^2 \sin(\phi-\theta), \tag{18}$$

$$2\frac{d\theta}{dt} = A^2 + 2\omega_0, \quad \frac{d\bar{x}}{dt} = k. \tag{19}$$

**3.2. Equilibrium states of the dynamical system (17)-(19).**

Equations (19) give rise to an FP, i.e., an equilibrium value of the soliton's amplitude and soliton's wavenumber:

$$A_*^2 = -2\omega_0 > 0, \; k_* = 0, \quad (20)$$

which exists for $\omega_0 < 0$. Then, it follows from Eq. (19) that the FP has $\theta = \text{const}$, and it is possible to set $\theta = 0$. Then, Eq. (18) yields the equilibrium value of the pump's phase shift:

$$\sin(\phi_*) = \frac{2\nu A_*}{\pi |f(\bar{x}_*)|}. \quad (21)$$

The existence of this value requires, obviously, that pump's strength at equilibrium point $\bar{x}_*$ must exceed a threshold value, $|f(\bar{x}_*)| \geq 2|\nu|A_*/\pi$. Then, equation (21) produces two values,

$$\phi_* = \arcsin\left(\frac{2\nu A_*}{|f(\bar{x}_*)|\pi}\right) + 2\pi p, \quad (22)$$

and

$$\phi_* = \pi - \arcsin\left(\frac{2\nu A_*}{|f(\bar{x}_*)|\pi}\right) + 2\pi p, \quad (23)$$

where $p$ is an integer. Straightforward linearization of Eqs. (17)-(19) around of the FPs readily shows that, in the framework of this dynamical system, for $\nu > 0$ (the linear loss in Eq. (1)) both equilibrium states are *unstable,* while for $\nu < 0$ (the linear amplification in Eq. (1), which may support the balance with the cubic dissipative pseudo-SRS term) the equilibrium state with phase (22) is *stable,* and one with phase (23) is *unstable*. For the case with zero loss/amplification ($\nu = 0$), the equilibrium state (20) with phase $\phi_* = 0$ is the respective phase plane is a center with frequency $\Omega = \sqrt{\pi |f(\bar{x}_*)|A_*/2}$. The equilibrium value of the soliton wavenumber $k_* = 0$ from Eq. (17) is achieved under condition

$$\frac{16\mu A_*^5}{45\pi} - \frac{d|f(\bar{x}_*)|}{dx}\cos(\phi_*) = 0. \quad (24)$$

For the stable equilibrium state with $\cos(\phi_*) > 0$, taking into account $\mu > 0$, we conclude that condition (24) holds for a positive gradient of the pumping amplitude, $d|f(\bar{x}_*)|/dx > 0$. With regard to condition $\cos(\phi_*) \leq 1$, relation (24) also imposes a condition on the magnitude of the gradient of the pump term at the equilibrium point: $d|f(\bar{x}_*)|/dx \geq 16\mu A_*^5/(45\pi)$. Then, taking into account Eqs. (21) and (22), condition (24) takes the form of

$$\frac{d|f(\bar{x}_*)|}{dx} = \frac{16\mu A_*^5}{45\pi\sqrt{1-\sin^2(\phi_*)}}.  \tag{25}$$

### 3.3. Time-dependent solutions of the system (17)-(19).

To produce an example of a nonstationary solution to Eqs. (17)-(19), we here consider a pump with the linear spatial profile $|f(x)| = f_0(1 + x/L)$ in Eq. (5) and choose parameters which make it possible to produce a characteristic solution:

$$\omega_0 = -1/2,\ \nu = -\pi/40,\ f_0 = 1/10,\ \mu = 1/20.  \tag{26}$$

In this case, the equilibrium value of the amplitude from Eq. (20) is $A_* = 1$. Substituting $\bar{x}_* = 0$ in expression (21) for the equilibrium phase, we obtain $\sin\varphi_* = -1/2$, hence the stable equilibrium has $\phi_* = -\pi/6$. In this case, the above-mentioned spatial scale $L_*$ of the inhomogeneous pump term is given by Eq. (25), taking into account that $d|f(\bar{x}_*)|/dx \equiv f_0/L_*$, as

$$L_* = 90\pi\sqrt{3}/32 \approx 15.  \tag{27}$$

Initial values for the wave number, coordinate, and phase are taken as

$$k(0) = k_* \equiv 0,\ \bar{x}(0) = \bar{x}_* = 0,\ \phi(0) = \phi_* \equiv -\pi/6,  \tag{28}$$

i.e., $\theta(0) = 0$.

For a relatively small deviation of the initial soliton's amplitude from the equilibrium value, viz., for

$$1/2 \equiv (1/2)A_* < A(0) < 2A_* \equiv 2,  \tag{29}$$

the numerical solution of Eqs. (17)-(19) drives $A(t)$ and $k(t)$ back to the FP, i.e., interval (29) is the *attraction basin* of this equilibrium state. A typical example of the respective dynamical regime is displayed in Figure 1(a), for initial conditions $A(0) = (3/2)A_* \equiv 3/2$.

For larger deviations of the initial soliton amplitude $A(0)$ from the equilibrium value $A_*$, viz., for $A(0) \geq 2A_* = 2$ or $A(0) \leq (1/2)A_* = 1/2$ (cf. Eq. (29), the numerical solution of Eqs. (17)-(19) demonstrates that the amplitude is growing, as shown in Figure 1(b) for the same parameter values as given by Eq. (26)-(28) and initial value $A(0) = 2A_* \equiv 2$, i.e., at the border of the attraction basin of the zero solution.

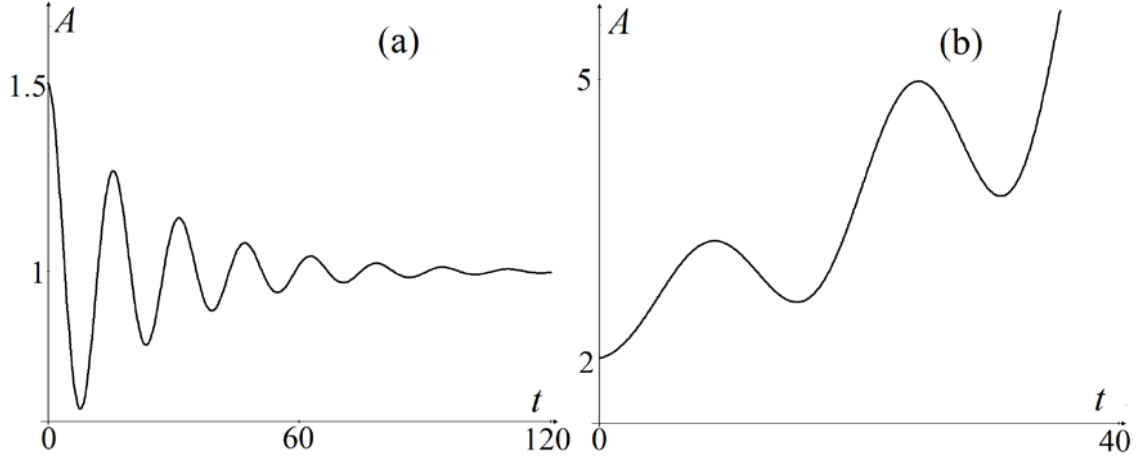

Fig. 1. A numerical solution of Eqs. (17)-(19) with parameters given by Eqs. (26)-(28) and different values of initial amplitude $A(0)$: (a) and (b) correspond to the initial amplitude $A(0) = (3/2)A_* \equiv 3/2$, and $A(0) = 2A_* \equiv 2$, respectively. The solution is represented by the amplitude as a function of time.

## 4. Results of full numerical simulations

### 4.1. LLE with a smooth localized pump

To check the above analytical predictions for the soliton dynamics, we simulated Eq. (4) with the the zero boundary conditions, the pump taken as per Eq. (5), and the input determined by Eq. (7),

$$U(x,0) = \tilde{V}_{bg}(x)\exp(i\phi(0)) + A(0)\text{sech}[A(0)x]\exp[ik(0)x], \qquad (30)$$

where $\tilde{V}_{bg}(x) = |f(x)|/2\omega_0$, $|f(x)| = f_0(1 + x/L)$, cf. Eq. (6). The parameters are taken here as in Eqs. (26), (27) and (28).

Simulations of Eq. (4) were performed for different initial amplitudes $A(0)$ in input (30). For $A(0) = A_* \equiv 1$ initial puls do not variables. Figure 2 display the results, severally, $A(0) = (3/2)A_* \equiv 3/2$ (a), and $A(0) = 2A_* \equiv 2$ (b), where $A_*$ is the respective equilibrium value (20) of system (17)-(19). In full agreement with the prediction of the approximation presented by Eqs. (17)-(19), these figures demonstrate, respectively, the unperturbed propagation of the soliton corresponding to the FP of the approximate equations, attraction of the non-equilibrium input to the FP in interval (28) of the initial values of the amplitude, and the growth of the amplitude outside of this interval. The numerical results for the time dependence of the spatial maximum of the absolute value of the field, $\max(|U(x,t)|)$, slightly differ (by about 5%) from the analytical results for the soliton`s amplitude $A(t)$ (Fig. 1). The difference is due to the presence of the background in the soliton region, $\tilde{V}_{bg} \approx 0,05$, which is about 5% of the amplitude.

Note that the stationary patterns produced by the simulations are stable, in spite of the presence of the linear amplification, accounted for by $\nu = -\pi/40$ in the parameter set (26). The action of the linear amplification on the vanishing background would make any localized state obviously unstable in a sufficiently large spatial domain, while here the stationary states are stable because of the finite size of the domain.

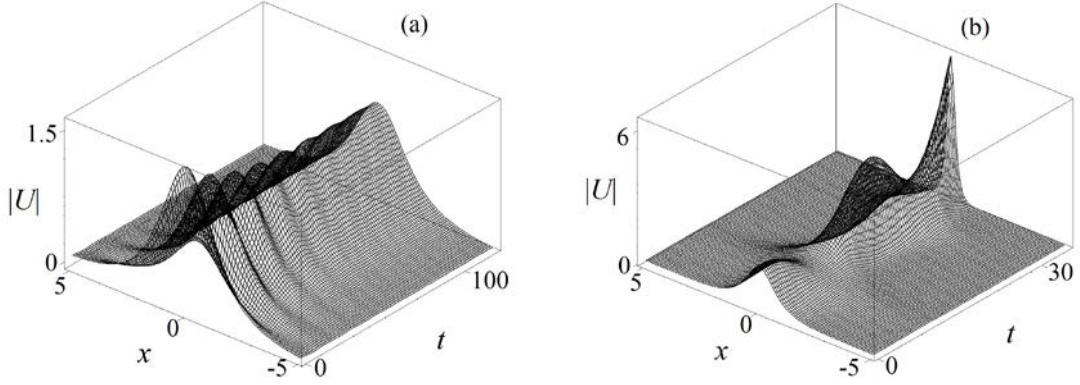

Fig. 2. Results of the simulations of Eq. (4) with parameter values (26)-(28) and input (30) with different initial amplitude $A(0)$ in Eq. (30): $A(0) = (3/2)A_* \equiv 3/2$ (a) and $A(0) = 2A_* \equiv 2$ (b).

### 4.2. LLE with a strongly localized spatial pump

For comparison with the above results obtained for the smooth spatially localized pump, we here aim to consider the case of a sharp (strongly localized) stationary pump term. To this end, we consider Eq. (4) with the pump in the form of a narrow Gaussian emulating the delta-function [61-63],

$$f(x) = f_0 \tilde{\delta}(x) \equiv f_0 \left(\sqrt{\pi}\varepsilon\right)^{-1} \exp\left(-x^2/\varepsilon^2\right), \tag{31}$$

zero initial condition, $U(x, t=0) = 0$ and parameters

$$\nu = 1/10, \ f_0 = 1/10, \ \mu = 1/10, \ \varepsilon = 1/10 \tag{32}$$

(note that, unlike the parameter set (26), the present one includes $\nu > 0$, i.e., the linear loss, rather than amplification). Figure 3 shows the spatial and temporal distribution of the wave-field modulus $|U(x,t)|$. It is seen that, in the course of the long evolution, the spatial distribution attains a stationary state with the spatial width $L_U \sim 10$.

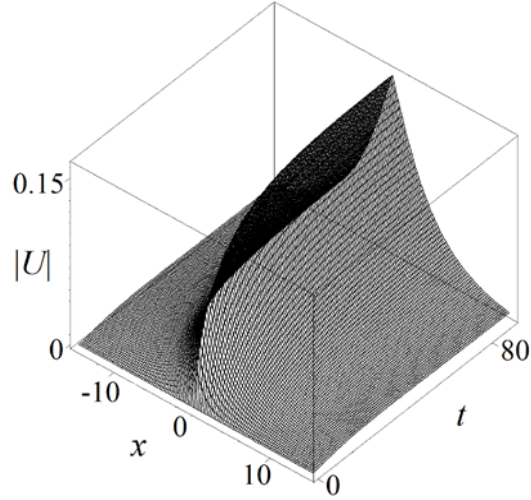

Fig. 3. Results of the simulations of Eq. (4) with the narrow pump taken as per Eq. (31), with parameters (32) and zero initial condition, $U(x, t=0)=0$.

## 5. Conclusion

We have considered the dynamics of the model which is based on the extended LLE (Lugiato-Lefever equation), which includes the cubic self-focusing, pseudo-SRS (simulated Raman scattering) nonlinear term, linear damping/amplification, and the external smoothly or sharply localized pump. The LLE can be derived from the ZS (Zakharov's system) whose HF (high-frequency) component is subject to the action of the linear damping and localized pump, while the equation for the LF (low-frequency) field includes effective viscosity. The model admits various physical realizations, such as a plasma pumped by external jets, coupled surface and internal waves in stratified fluids, etc. The extended LLE gives rise to the quasi-soliton state with a nonzero background. These solutions are predicted, by means of the analytical approximation, as the FP (fixed point) of the system of evolution equations for the HF norm, momentum, and center-of-mass coordinate. The FP's attraction basin is identified too. The analytical prediction is corroborated by direct simulations of the underlying LLE. In addition to that, a quasi-singular mode, supported by the pump in the form of the delta-function, is found too.

## Acknowledgment

This work was supported, in part, by Israel Science Foundation through grant No. 1286/17.